\documentclass{aa}
\usepackage{graphicx}
\usepackage{txfonts} 
\usepackage{hyperref}
\graphicspath{{./}{Figures/}}

\begin{document}

\title{JWST reveals a high fraction of disk breaks at $1\leq z\leq 3$}

   \author{Dewang Xu\inst{\ref{PKU},}\inst{\ref{KIAA}}
            \and
               Si-Yue Yu\inst{\ref{mpifr}}\thanks{Corresponding author}
          }

   \institute{ 
   Department of Astronomy, Peking University, 5 Yiheyuan Road, Haidian District, Beijing, 100871, China \label{PKU}
   \and
   The Kavli Institute for Astronomy and Astrophysics, Peking University, 5 Yiheyuan Road, Haidian District, Beijing, 100871, China\label{KIAA}
   \and
    Max-Planck-Institut für Radioastronomie, Auf dem Hügel 69, 53121 Bonn, Germany  \\ \email{phyyueyu@gmail.com, syu@mpifr-bonn.mpg.de}\label{mpifr} 
            }

\abstract{
We analyzed the deconvolved surface brightness profiles of 247 massive and angularly large disk galaxies at $1\leq z\leq 3$ to study high-redshift disk breaks, using F356W-band images from the Cosmic Evolution Early Release Science survey (CEERS). We found that 12.6\% of these galaxies exhibit type~I (exponential) profiles, 56.7\% exhibit type~II (down-bending) profiles, and 34.8\% exhibit type~III (up-bending) profiles. Moreover, we showed that galaxies that are more massive, centrally concentrated, or redder, tend to show fewer type~II and more type~III breaks. These fractions and the detected dependencies on galaxy properties are in good agreement with those observed in the Local Universe. In particular, the ratio of the type~II disk break radius to the bar radius in barred galaxies typically peaks at a value of 2.25, perhaps due to bar-induced radial migration. However, the timescale for secular evolution may be too lengthy to explain the observed breaks at such high redshifts. Instead, violent disk instabilities may be responsible, where spiral arms and clumps torque fling out the material, leading to the formation of outer exponential disks. Our results provide further evidence for the assertion that the Hubble Sequence was already in place during these early periods.
}

\maketitle

\section{Introduction} \label{sec:intro}
The radial distribution of stars in galaxy disks has been widely described by an exponential function \citep{deVaucouleurs:1958, deVaucouleurs:1959b,freeman_1970} that may be a result of the redistribution of mass and angular momentum due to viscosity \citep{lin_prigles_1987}. 
Although an exponential function provides on average a reasonably successful description, numerous recent works have pointed out that the outer regions of many galaxies deviate from this simple functional form and could instead be better parameterized by a double exponential function \citep{Erwin:2005, pohlen2006, Erwin:2008, Gutirrez_2011, Laine_2014, watkins_2019, tang2020}.

Following the nomenclature of \citet{freeman_1970}, which was further developed by \citet{pohlen2006} and \citet{Erwin:2008}, the disk surface brightness profiles are divided primarily into three types: type~I, which aptly follows a single exponential function; type~II, which is best fitted by a double exponential function with a steeper slope in the outer profile; and type~III, best described by a double exponential function with a shallower slope in the outer profile. There are approximately 70\%\textendash80\% nearby disk galaxies having type~II or III disk breaks \citep{pohlen2006, Erwin:2008, Gutirrez_2011, Laine_2014, tang2020}. It has been found that type~II disks are more prevalent in late-type galaxies, while type~III disk breaks occur more frequently in early-type systems \citep{pohlen2006, Erwin:2008, Gutirrez_2011, laine_2016, tang2020}. Furthermore, the fraction of type~II profiles present in cluster disk galaxies is suppressed in comparison to those found in field galaxies (\citealt{Erwin2012}; \citealt{Roediger2012}; \citealt{Raj2019}; \citealt{Pfeffer2022}; but see \citealt{Pranger2017}). Type~II disk are likely caused by a threshold in molecular surface density for forming new stars \citep{Kennicutt1989, martin2001, Schaye_2004, Elmegreen2006}, radial migration driven by bars or spirals \citep{sellwood2002, Debattista2006, Bakos2008, Roskar2008, Minchev2012, dimatteo2013}, or disk instabilities in the earlier Universe \citep{Bournaud2007}. In contrast, relatively little is known about the origins of type~III disks. Proposed scenarios have included mergers \citep{Bekki1998}, tidal disturbances \citep[e.g.,][]{watkins_2019}, confusion with a stellar halo \citep{Martin2012, Martin2014, Peters2017}, and in situ star formation due to gas accretion \citep{wang:2018}.

Thanks to the {\it Hubble} Space Telescope (HST), galaxies exhibiting disk breaks at intermediate redshifts ($z\lesssim 1$) have been found \citep{Perez2004, Trujillo2005, Azzollini2008, Bakos2008, Borlaff2018}. However, the exploration of disk breaks at higher redshifts ($z\gtrsim 1$) using HST has been limited, due to the clumpy nature of high-redshift galaxies observed by HST \citep{Conselice2008, Mortlock2013}. Recent studies, leveraging the unprecedented sensitivity and resolution of the {\it James Webb} Space Telescope (JWST) in the infrared, have revealed that a substantial fraction of high-redshift galaxies are regular disks, rather than being predominantly clumpy \citep{Ferreira2022a, Ferreira2022b, Kartaltepe2023}. These galaxies display features such as spirals and bars, as seen in nearby galaxies \citep{Chen2022, Fudamoto2022, Guo2023, LeConte:2023}.  These findings raise new questions about the occurrence and frequency of disk breaks in galaxies at high redshifts. These answers may contribute to our understanding of galaxy evolution. To investigate these questions further in this work, we analyze the surface brightness profiles of disk galaxies observed by JWST at redshifts $1\leq z\leq 3$.

\section{Data and sample} \label{data_jwst}
We analyzed the galaxies observed in the Cosmic Evolution Early Release Science (CEERS) Survey (PI: Finkelstein, ID=1345, \citealt{Finkelstein:2022}). The sample selection criteria were based on redshift, rest-frame color, and stellar mass from the 3D-HST catalog \citep{Brammer2012, Skelton2014}. We focused on galaxies with redshifts of $1 \leq z \leq 3$ and stellar masses  of $M_*\geq 10^{10} M_{\odot}$. The diagram of rest-frame $V-J$ versus rest-frame $U-V$, along with the demarcation line proposed by \cite{Williams2009}, were used to identify star-forming galaxies, most of which are disk galaxies.  We utilized the mosaic image product provided by the DAWN JWST Archive \citep{Valentino2023}\footnote{https://dawn-cph.github.io/dja; version 7 is used.}. We use \texttt{WebbPSF} \citep[version 1.2.1, ][]{Perrin:2014} to construct the point spread function (PSF). Specifically, this study employs the F356W-band mosaic images for their effectiveness in tracing older stellar populations, whereas the F444W band is avoided due to its broader PSF that tends to smooth out the disk structure further. This uniform use of the F356W band also minimizes uncertainties associated with deconvolution.  Galaxies that are spheroidal in shape, those located at the edges of the mosaic image, and those affected by significant light contamination from nearby sources are excluded upon visual inspection. We measured the axis ratio ($b/a$) and half-light radius ($R_{50}$) for each galaxy and then further restricted our selection to galaxies with $b/a\geq 0.5$ to avoid severe projection effects and those with $R_{50}\geq 2\times \text{FWHM}$ for obtaining accurate deconvolved surface brightness profiles. The deconvolution is elaborated in the following section. Our sample selection results in a final sample of  247 galaxies. 

\begin{figure*}
     \centering
        \includegraphics[width=1\linewidth]{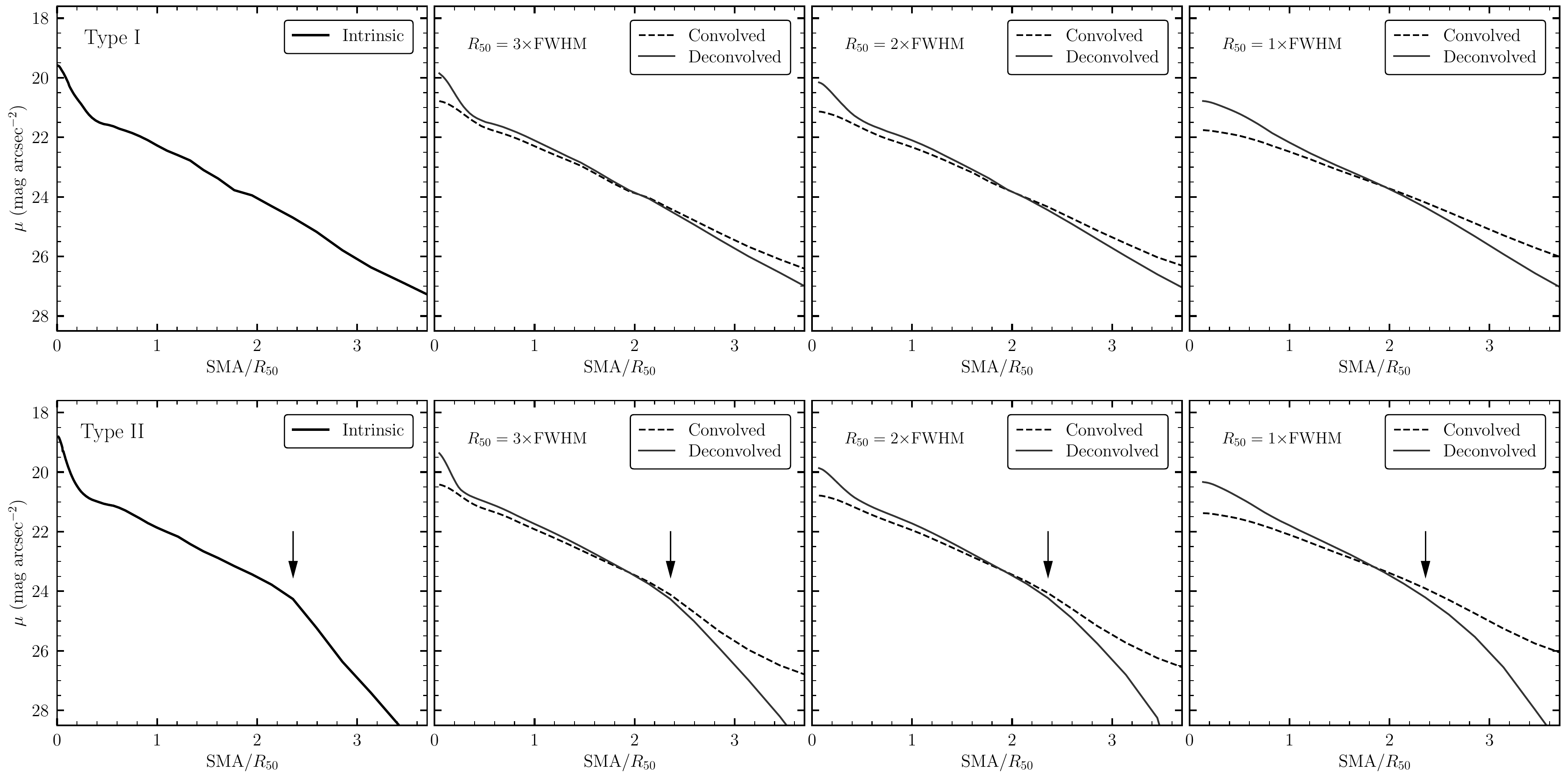}
     \caption{Illustration of the deconvolved surface brightness profiles. The left two panels illustrate the intrinsic profiles obtained from high-resolution images of nearby galaxies, featuring ESO~112-005 at the top and NGC~12 at the bottom, respectively. The panels in the subsequent columns present the profiles extracted from artificial downsized images convolved with the JWST F356W PSF, indicated by dashed curves, and present the deconvolved profiles, shown as solid curves. The resolution of the artificial image is indicated by the ratio of galaxy half-light radius ($R_{50}$) to PSF FWHM, denoted at the top of the panel. The arrow denotes the disk break seen in the intrinsic profile.
     }
     \label{fig:deconv} 
\end{figure*}

\begin{figure*}
     \centering
        \includegraphics[width=0.65\linewidth]{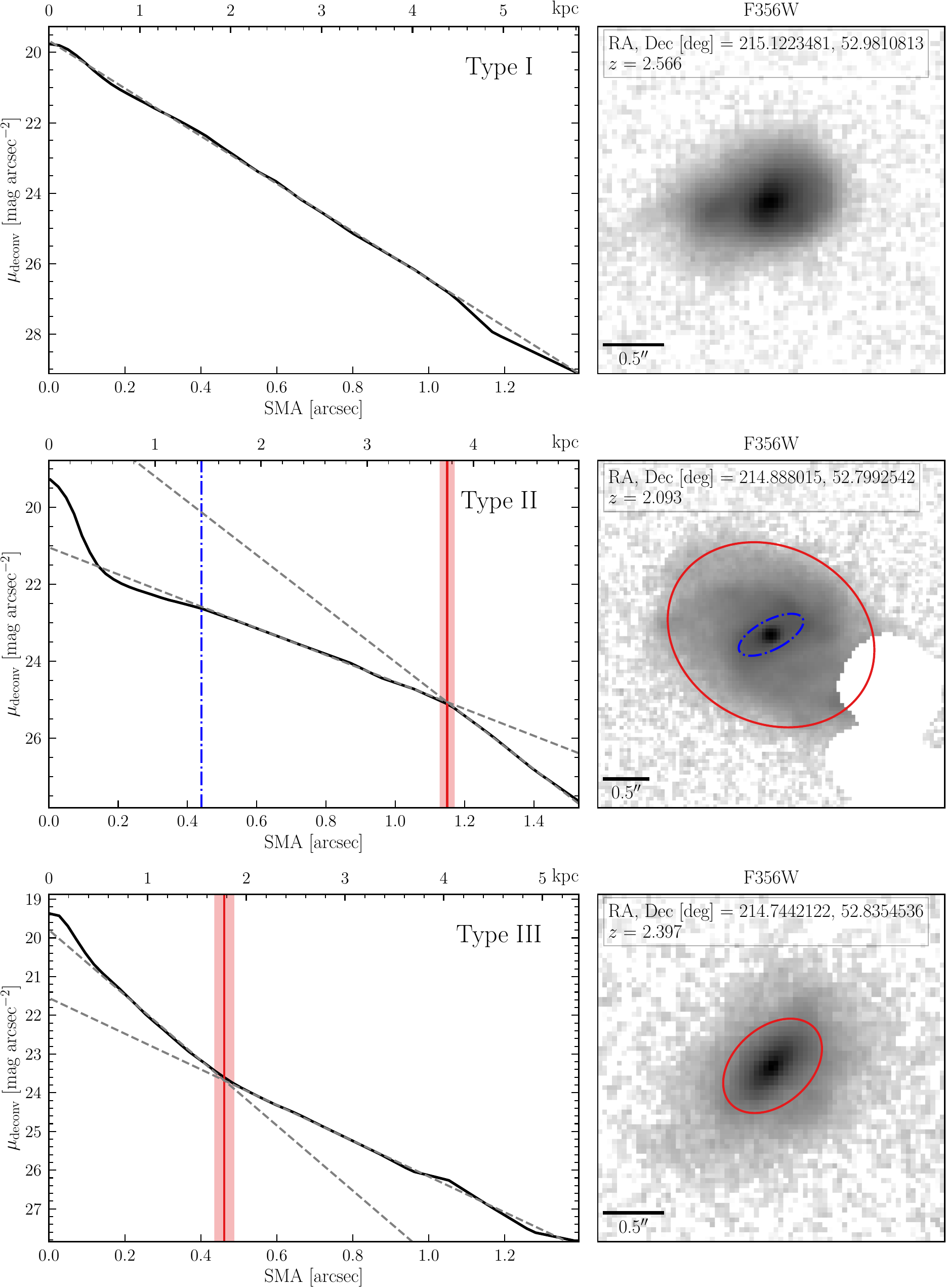}
     \caption{Classification of the deconvolved disk profiles into three types: type~I (exponential; top), type~II (down-bending; middle), and type~III (up-bending; bottom). The left panels plot the deconvolved surface brightness profiles as a function of semi-major axis (SMA) extracted from deconvolved images. In these, the blue dashed-point line marks the measured bar radius (if present), and the solid red line indicates the identified disk break radius. The dashed lines represent the best-fit exponential functions. On the right, the panels display the deconvolved F356W-band images, with overplotted ellipses corresponding to the bar radius (if present) and disk break radius in the same line styles as in the left panels.
     }
     \label{fig:db_example} 
\end{figure*}

\section{Deconvolution of surface brightness profiles and classification of disk breaks}
As galaxies at higher redshifts become angularly smaller due to longer cosmological distance or intrinsic size evolution, their structures become more blurred because of the PSF smoothing effect \citep[e.g.,][]{Liang2023}. This PSF effect leads to a flattening of the observed surface brightness profile, especially in galaxies at high redshifts. Therefore, deconvolution is essential to minimize the PSF smoothing effect and recover the intrinsic surface brightness profiles of these galaxies. The Richardson-Lucy algorithm \citep{Richardson:1972, Lucy:1974}, a popular deconvolution method, can severely amplify noise and introduce  artificial flux in low surface brightness regions, making it unsuitable for studying surface brightness profiles. Alternatively, we deconvolved galaxy images throughout a 2D model fitting process, adapting the strategy used in \cite{Trujillo:2016}. For each galaxy image, we use {\tt sep} \citep{Barbary2016} to generate a mask to avoid contamination from other sources. By involving the F356W PSF, we use \texttt{imcascade}, developed by  \cite{Miller:2021}, to fit the galaxy using their multi-Gaussian expansion method to obtain the intrinsic galaxy model that has no PSF effects. The deconvolved image is obtained by subtracting the intrinsic model convolved with the PSF from the original image and then adding the intrinsic model back to the residual.  This process effectively minimizes the contribution of the PSF, facilitating the extraction of deconvolved surface brightness profiles, yet it cannot deconvolve finer local structures, such as spiral arms. The center from the fitting is taken as the galaxy center, while the $b/a$ and position angle (PA) are measured on the outskirts of the galaxy using isophotes. With fixed galaxy center, $b/a$, and PA, we use \texttt{photutils} \citep{Bradley:2020photutils} to extract azimuthally averaged surface brightness profiles from the deconvolved images. 

In Fig.~\ref{fig:deconv}, we illustrate the effectiveness and the limitations of the deconvolution process using the {\it r}-band images of ESO~112-005, which exhibits a type~I profile (top), and NGC~12, showcasing a type~II profile (bottom), as cleaned of stars and provided by \cite{Yu2023}. The left panels show the intrinsic profiles extracted from the high-resolution stars-cleaned images. The panels in the subsequent columns illustrate the profiles before and after deconvolution, represented by dashed and solid curves, respectively. These comparisons are made using simulated images with various ratios of $R_{50}$ to the F356W FWHM. The effect of convolution tends to smooth the surface brightness profiles, transforming them into apparent type~I profiles. The dashed curves in the bottom panels illustrate how, in the absence of deconvolution, an intrinsic type II profile could be mistakenly identified as a type I profile. When $R_{50}\geq2\times \text{FWHM}$, the deconvolved profiles closely resemble the intrinsic profile, including a similar disk break. In contrast, for $R_{50}\leq 1\times \text{FWHM}$, the deconvolved profile is highly non-exponential, with a slope that changes across the entire radial extent, and thus fails to reconstruct the intrinsic profile with its characteristic break. Therefore, the deconvolution is most effective when $R_{50}\geq2\times \text{FWHM}$, a criterion that has been adopted for sample selection in Sect.~\ref{data_jwst}.

We identified bars by searching local peaks in the isophotal ellipcity profile \citep[e.g.,][]{Erwin2003, Yu2022, Liang2023}, leading to a bar fraction of 42\%, and then measured their bar radius ($R_{\rm bar}$). The ellipticity-drop method is deprecated for high-redshift galaxies due to the smoothing of drops into peaks by PSF \citep{Liang2023}. Following the classification scheme of \citet{pohlen2006} and \citet{Erwin:2008}, we classify the surface brightness profiles using three main types: type~I (exponential), type~II (down-bending), and type~III (up-bending), with a combination of type~II and III (mixed types) in case of multiple breaks. The process involved manually determining the disk regions above the surface brightness limit, generally above 28 mag\,arcsec$^{-2}$, and fitting one or more exponential functions to the profile; disk breaks were identified if the scale length ratio of two adjacent exponentials was greater than 1.1 or less than 0.9 \citep{wang:2018}, and the break radius ($R_{\rm brk}$) was determined at the intersection point of these exponentials, with its uncertainty estimated by assigning a 20\% uncertainty to the selected regions and re-fitting. In barred galaxies, type~II profiles are further divided into inner break type~II.i and outer break type~II.o, depending on whether the disk break radius is beyond the bar radius. For simplicity, we classify the type~II.o profiles as type~II, the same class as the type~II profiles in unbarred galaxies.  Examples of three CEERS disk galaxies at $z>2$, showcasing type~I, II, and III profiles are illustrated in the top, middle, and bottom panels of Fig.~\ref{fig:db_example}.

\begin{figure*}
     \centering
        \includegraphics[width=1\linewidth]{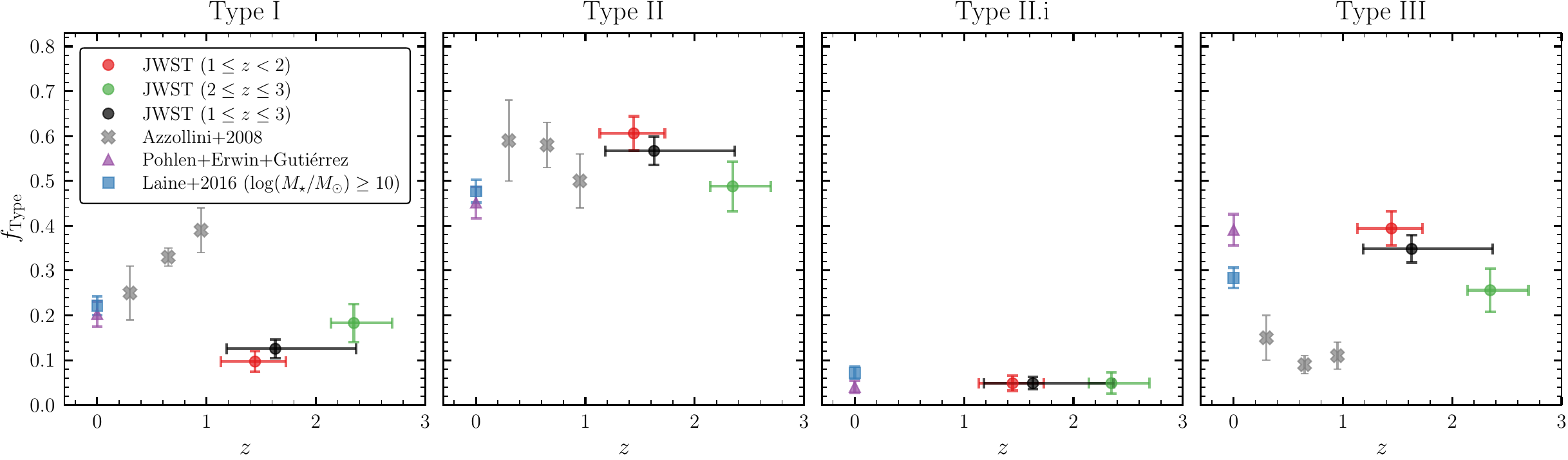}
     \caption{Comparison of our derived fractions of type I, type II, type II.i, and type III profiles with those in previous studies.  Our results are calculated in three redshift bins and are marked by points positioned at the median redshift. Results derived from the combined data of studies \cite{pohlen2006}, \cite{Erwin:2008}, and \cite{Gutirrez_2011} in the same series are marked as purple triangles; those reported in \cite{laine_2016} are marked by blue squares. Results based on HST reported in \cite{Azzollini2008} are shown by grey crosses. Error bars associated with the symbols denote the statistical uncertainties.
     }
     \label{fig:frac_comp} 
\end{figure*}

\section{Results}
\subsection{Statistics of disk profile types}
For the whole sample within $1\leq z\leq 3$, our analysis reveals that the fraction of type~I disks is $f_{\rm I}=12.6\%$, the fraction of type~II disks is $f_{\rm II}=56.7\%$, the fraction of type~II.i disks is $f_{\rm II.i}=4.9\%$, and the fraction of type~III disk is $f_{\rm III}=34.8\%$. These fractions do not change significantly when the sample is divided into redshift bins of $1\leq z<2$ and $2 \leq z \leq 3$.  Our derived fractions, together with those reported in previous studies, are plotted as a function of redshift in Fig.~\ref{fig:frac_comp}. The fractions derived from optical images of nearby galaxies, calculated on the basis of on an aggregation of data from several studies within the same series, specifically those conducted by \cite{pohlen2006}, \cite{Erwin:2008}, and \cite{Gutirrez_2011}, are represented by purple triangles. We also include classification results derived from NIR images of nearby galaxies, as reported by \cite{laine_2016}, but only focus on galaxies with $\log M_*/M_{\odot}\geq 10$, same with the mass cut used in this study. The fractions of profile types calculated in this manner are denoted by blue squares. These galaxies span a wide range of environment from field to cluster \citep{laine_2016}. Our derived fractions for high-redshift CEERS galaxies are broadly consistent with those obtained from optical image or NIR images of nearby galaxies. Nevertheless, we note that there is a minor increase in $f_{\rm II}$ or a decrease in $f_{\rm I}$, which may be caused by minor cosmic evolution or the difference in the sample properties. Compared to the findings obtained from intermediate-redshift ($z\leq1$) galaxies reported by \cite{Azzollini2008}, which are marked with grey crosses, our $f_{\mathrm{III}}$ values are higher. This difference is likely attributed to the classification approach we adopted from \cite{Erwin:2008}, particularly in how we handle mixed-type profiles. While \cite{Azzollini2008} tended to classify these mixed types into a single type~II type, our methodology allows for a more detailed classification, leading to discrepancies between our results and those reported by \cite{Azzollini2008}.

\subsection{Dependence of disk profile types on physical properties}
\begin{figure*}
     \centering
        \includegraphics[width=1\linewidth]{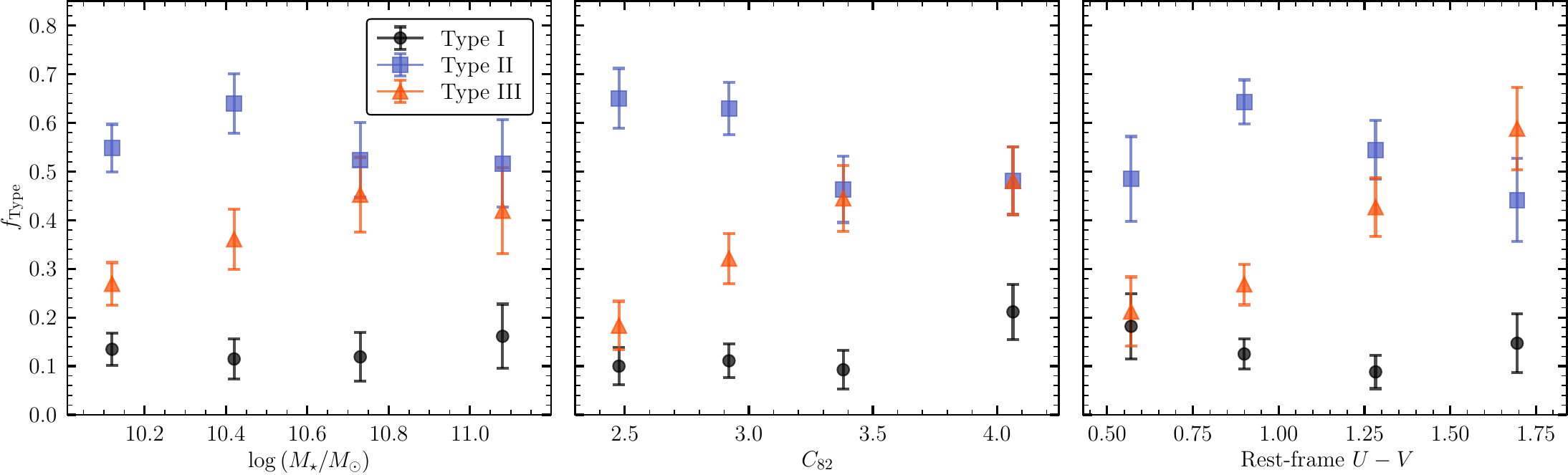}
     \caption{Dependence of the factions of profile types on stellar mass (left), concentration index $C_{82}$ (middle), and rest-frame color $U-V$ (right). 
     }
     \label{fig:frac_dependence} 
\end{figure*}

We explored the dependence of the fractions of disk profile types on stellar mass, concentration index, and rest-frame $U-V$ color in Fig.~\ref{fig:frac_dependence}. We found the most pronounced correlation with the concentration index. As the concentration index increases, the fraction of type~II decreases while the fraction of type~III increases, and the fraction of type~I remains nearly unchanged. Such a strong dependence for type~III fraction on stellar concentration is in good agreement with that for nearby galaxies, as shown in \cite{wang:2018}. Although the correlation with stellar mass and that with color index are not as pronounced, there is a weak trend that more massive or redder galaxies tend to show lower type~II fractions and higher type~III fractions. There was no clear trend found for type~I profiles. Our results are consistent with previous studies on nearby galaxies \citep[e.g., ][]{Gutirrez_2011, tang2020}, which showed that type~III profiles are more prevalent in early-type galaxies, while type~II profiles are more common in late-types. 

\subsection{Connection between break radius with bar radius and disk size}

\begin{figure}
     \centering
        \includegraphics[width=1\linewidth]{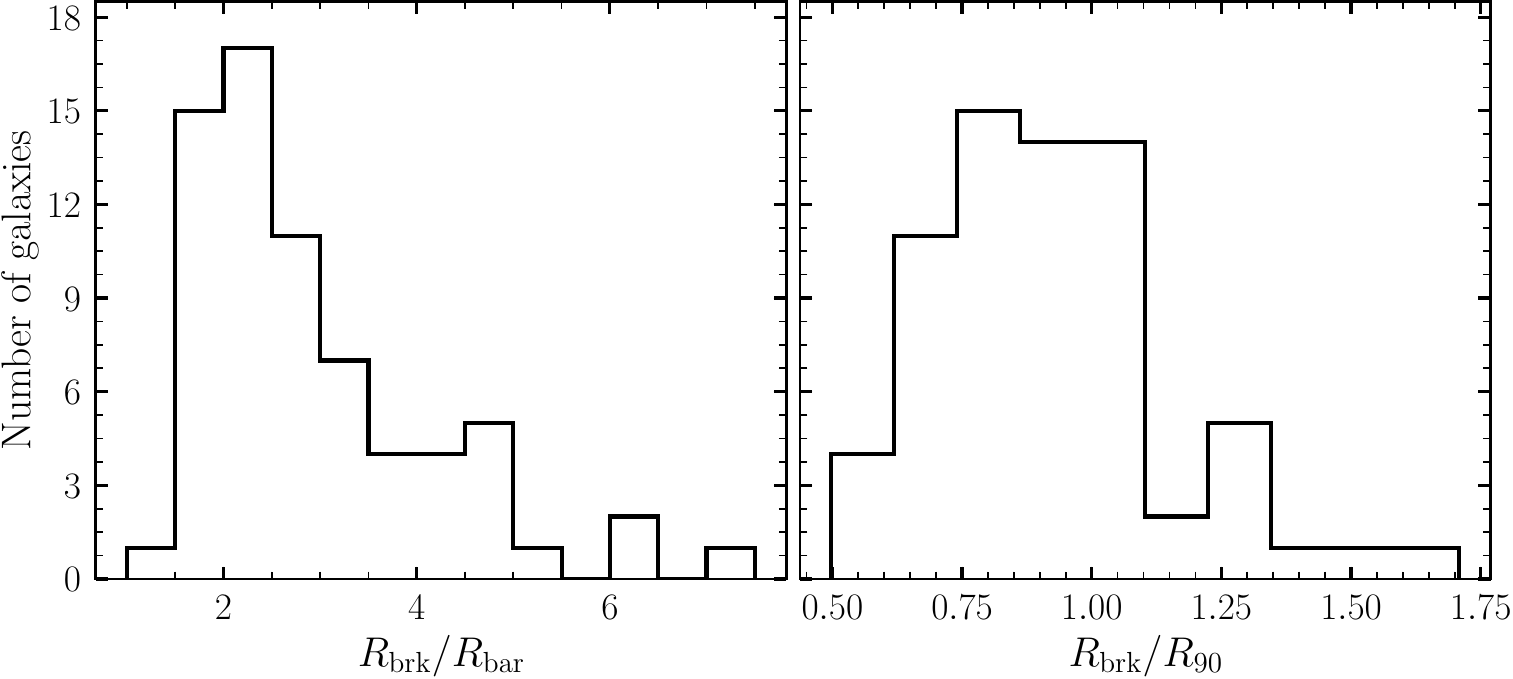}
     \caption{Distribution of the ratio of type~II break radius ($R_{\rm brk}$) to bar radius ($R_{\rm bar}$) on the left and the ratio of $R_{\rm brk}$ to $R_{90}$ on the right.
     }
     \label{fig:bar_stats} 
\end{figure}

In Fig.~\ref{fig:bar_stats}, we present the distribution of the ratio of type~II break radius ($R_{\rm brk}$) to bar radius ($R_{\rm bar}$) on the left, and the ratio of $R_{\rm brk}$ to $R_{90}$ (a measure of disk size that encloses 90\% of galaxy flux) on the right. We show that the distribution of $R_{\rm brk}/R_{\rm bar}$ ratios peaks at approximately 2.25, displaying an asymmetric shape with a skewness towards higher values; the distribution of $R_{\rm brk}/R_{\rm 90}$ ratios peak at approximately 0.8. To estimate which of these two quantities exhibits a stronger correlation with $R_{\rm brk}$, we compute the Pearson correlation coefficient ($\rho$) for both. This yields $\rho(R_{\rm brk}, R_{\rm bar})=0.52$ and $\rho(R_{\rm brk}, R_{90})=0.66$. Therefore, there is no significant evidence to suggest that the break radius is more closely related to the bar radius than it is to the disk size.

\section{Discussion and conclusion}
A significant proportion of nearby disk galaxies exhibit breaks in their surface brightness profiles, mainly categorized as type~I, type~II, and type~III. Studying the potential cosmic evolution of the fraction of these disk break types could provide insights into the origin of disk breaks and the process of galaxy evolution. Through the analysis of deconvolved surface brightness profiles of 247 CEERS disk galaxies at $1\leq z\leq 3$, we found that not only are their fraction of break types comparable with those observed in the Local Universe \citep{pohlen2006, Erwin:2008, Gutirrez_2011, Laine_2014}, but that the dependence of fractions on galaxy properties is similar to previous results based on nearby galaxies \citep{Gutirrez_2011, wang:2018, tang2020}.  This consistency implies that despite galaxies being intrinsically more luminous \citep[e.g.,][]{Yu2023} and more compact \citep[e.g.,][]{vanderWel2014} at the earlier cosmic time, the phenomenon of disk breaks had already been firmly established in the early Universe. Moreover, our findings are in good agreement with recent observations by the JWST, which reveal that regular stellar structures such as bars and spirals, closely resembling their nearby counterparts, are present in these high-redshift galaxies  \citep[e.g.,][]{Chen2022, Fudamoto2022, Guo2023}. 

Environment may play a role in sculpting the surface brightness profiles. Clusters are rarer at higher redshifts \citep{Gladders2005, Pacaud2007, Hasselfield2013}, reducing the likelihood of type~II profile suppression within cluster environments  \citep{Erwin2012, Roediger2012, Raj2019, Pfeffer2022}. This potentially contributes to an increase in the fraction of type~II profiles at higher redshifts and leads to the observed modest elevation in our reported type~II fraction compared to those identified in the Local Universe (Fig.~\ref{fig:frac_comp}). In contrast, given that type~III profiles are more frequent in denser environment \citep{watkins_2019, Pfeffer2022}, the type~III fraction at higher redshifts may decrease. Nonetheless, the increased merger rates \citep{Rodriguez-Gomez2015} and gas fractions \citep{Geach2011, Magdis2012} observed at higher redshifts may promote the formation of type~III profiles through mergers \citep{Bekki1998} and in situ star formation fuelled by gas accretion \citep{wang:2018}, respectively, compensating for the aforementioned decline.

Previous observational studies have suggested a connection of type~II disk break to the outer Lindblad resonance (OLR) of the bar structure \citep[e.g., ][]{pohlen2006, Erwin:2008, Munoz-Mateos2013, Laine_2014}. Simulations have shown that bars are effective drivers of radial migration through their corotation resonance (CR), making stars travel several kiloparsecs radially both inwards and outwards, leading to the formation of a down-bending break roughly at the OLR \citep[e.g.,][]{Debattista2006, Minchev2012}. Given the position of the OLR being about two to three times the bar radius \citep{Erwin:2008, Munoz-Mateos2013}, the bar-driven radial migration leads to $R_{\rm brk}/R_{\rm bar}\approx$2\textendash3 (on average), which could potentially explain our reported distribution of $R_{\rm brk}/R_{\rm bar}$. The observation of this peak from the high-redshift Universe ($1\leq z\leq 3$), as reported in this study, persisting into the Local Universe ($z\approx0$), which was explored in \cite{Munoz-Mateos2013}, implies that these disk breaks are likely long-lived structures and evolve in conjunction with the bars. For unbarred galaxies, type~II breaks could be a result of radial migration driven solely by spirals \citep{Martinez-Bautista2021}, particularly in grand-design spiral galaxies with density-wave arms.

However, we note that the distribution of $R_{\rm brk}/R_{\rm bar}$ spans a wide range and that the breaks do not show a stronger correlation with bar than disk, thereby suggesting that bars might not predominantly influence the observed patterns. Furthermore, since the scenario of bar-driven radial migration operates over a few Gyr \citep{Roskar2008, Minchev2012}, the disks at $z=3$ may not have time to experience such secular evolution processes. Rather, the dynamical evolution of high-redshifts disks could be primarily governed by more violent disk instabilities \citep[e.g.,][]{Dekel2009}. This process is fast and can rapidly generate type~II disk breaks by flinging out materials through spiral arms and clump torques in less than 1\,Gyr \citep{Bournaud2007}.

Our study reveals a high fraction of disk breaks in galaxies at redshifts $1 \leq z \leq 3$ in the JWST CEERS field, supporting the findings of  \cite{Ferreira2022b} and \cite{Kartaltepe2023}, which state that the Hubble sequence was already in place at these higher redshifts. While the exact mechanisms behind the formation of these breaks remain uncertain, future studies focusing on their color profiles could provide deeper insights into their origins. Such investigations may be crucial for understanding the interplay between galaxy structure and evolution across cosmic time, potentially revealing how high-redshift galaxies evolved into the structures we now observe in the Local Universe.

\begin{acknowledgements}
We thank the referee for his insightful and constructive feedback, which significantly enhanced the quality and clarity of this letter. We thank the fruitful discussion with Cheng Cheng. SYY thanks Prof. Karl Menten for his support. This work is based on observations taken by the 3D-HST Treasury Program (GO 12177 and 12328) with the NASA/ESA HST, which is operated by the Association of Universities for Research in Astronomy, Inc., under NASA contract NAS5-26555. (Some of) The data products presented herein were retrieved from the Dawn JWST Archive (DJA). DJA is an initiative of the Cosmic Dawn Center, which is funded by the Danish National Research Foundation under grant No. 140.

\end{acknowledgements}

\bibliographystyle{aa}

\begin{thebibliography}{72}
\expandafter\ifx\csname natexlab\endcsname\relax\def\natexlab#1{#1}\fi

\bibitem[{{Azzollini} {et~al.}(2008){Azzollini}, {Trujillo}, \&
  {Beckman}}]{Azzollini2008}
{Azzollini}, R., {Trujillo}, I., \& {Beckman}, J.~E. 2008, \apj, 684, 1026

\bibitem[{{Bakos} {et~al.}(2008){Bakos}, {Trujillo}, \& {Pohlen}}]{Bakos2008}
{Bakos}, J., {Trujillo}, I., \& {Pohlen}, M. 2008, \apjl, 683, L103

\bibitem[{Barbary(2016)}]{Barbary2016}
Barbary, K. 2016, Journal of Open Source Software, 1, 58

\bibitem[{{Bekki}(1998)}]{Bekki1998}
{Bekki}, K. 1998, \apjl, 502, L133

\bibitem[{{Borlaff} {et~al.}(2018){Borlaff}, {Eliche-Moral}, {Beckman},
  {Vazdekis}, {Lumbreras-Calle}, {Ciambur}, {P{\'e}rez-Gonz{\'a}lez},
  {Cardiel}, {Barro}, \& {Cava}}]{Borlaff2018}
{Borlaff}, A., {Eliche-Moral}, M.~C., {Beckman}, J.~E., {et~al.} 2018, \aap,
  615, A26

\bibitem[{{Bournaud} {et~al.}(2007){Bournaud}, {Elmegreen}, \&
  {Elmegreen}}]{Bournaud2007}
{Bournaud}, F., {Elmegreen}, B.~G., \& {Elmegreen}, D.~M. 2007, \apj, 670, 237

\bibitem[{{Bradley} {et~al.}(2020){Bradley}, {Sip{\H{o}}cz}, {Robitaille},
  {Tollerud}, {Vin{\'\i}cius}, {Deil}, {Barbary}, {Wilson}, {Busko},
  {G{\"u}nther}, {Cara}, {Conseil}, {Bostroem}, {Droettboom}, {Bray}, {Andersen
  Bratholm}, {Lim}, {Barentsen}, {Craig}, {Pascual}, {Perren}, {Greco},
  {Donath}, {De Val-Borro}, {Kerzendorf}, {Bach}, {Weaver}, {D'Eugenio},
  {Souchereau}, \& {Ferreira}}]{Bradley:2020photutils}
{Bradley}, L., {Sip{\H{o}}cz}, B., {Robitaille}, T., {et~al.} 2020,
  {astropy/photutils: 1.0.0}

\bibitem[{{Brammer} {et~al.}(2012){Brammer}, {van Dokkum}, {Franx},
  {Fumagalli}, {Patel}, {Rix}, {Skelton}, {Kriek}, {Nelson}, {Schmidt},
  {Bezanson}, {da Cunha}, {Erb}, {Fan}, {F{\"o}rster Schreiber}, {Illingworth},
  {Labb{\'e}}, {Leja}, {Lundgren}, {Magee}, {Marchesini}, {McCarthy},
  {Momcheva}, {Muzzin}, {Quadri}, {Steidel}, {Tal}, {Wake}, {Whitaker}, \&
  {Williams}}]{Brammer2012}
{Brammer}, G.~B., {van Dokkum}, P.~G., {Franx}, M., {et~al.} 2012, \apjs, 200,
  13

\bibitem[{{Chen} {et~al.}(2022){Chen}, {Gao}, {Hsu}, {Liao}, {Ling}, {Lo},
  {Smail}, {Wang}, \& {Wang}}]{Chen2022}
{Chen}, C.-C., {Gao}, Z.-K., {Hsu}, Q.-N., {et~al.} 2022, \apjl, 939, L7

\bibitem[{{Conselice} {et~al.}(2008){Conselice}, {Rajgor}, \&
  {Myers}}]{Conselice2008}
{Conselice}, C.~J., {Rajgor}, S., \& {Myers}, R. 2008, \mnras, 386, 909

\bibitem[{{de Vaucouleurs}(1958)}]{deVaucouleurs:1958}
{de Vaucouleurs}, G. 1958, \apj, 128, 465

\bibitem[{{de Vaucouleurs}(1959)}]{deVaucouleurs:1959b}
{de Vaucouleurs}, G. 1959, Handbuch der Physik, 53, 311

\bibitem[{{Debattista} {et~al.}(2006){Debattista}, {Mayer}, {Carollo}, {Moore},
  {Wadsley}, \& {Quinn}}]{Debattista2006}
{Debattista}, V.~P., {Mayer}, L., {Carollo}, C.~M., {et~al.} 2006, \apj, 645,
  209

\bibitem[{{Dekel} {et~al.}(2009){Dekel}, {Sari}, \& {Ceverino}}]{Dekel2009}
{Dekel}, A., {Sari}, R., \& {Ceverino}, D. 2009, \apj, 703, 785

\bibitem[{{Di Matteo} {et~al.}(2013){Di Matteo}, {Haywood}, {Combes},
  {Semelin}, \& {Snaith}}]{dimatteo2013}
{Di Matteo}, P., {Haywood}, M., {Combes}, F., {Semelin}, B., \& {Snaith}, O.~N.
  2013, \aap, 553, A102

\bibitem[{{Elmegreen} \& {Hunter}(2006)}]{Elmegreen2006}
{Elmegreen}, B.~G. \& {Hunter}, D.~A. 2006, \apj, 636, 712

\bibitem[{{Erwin} {et~al.}(2005){Erwin}, {Beckman}, \& {Pohlen}}]{Erwin:2005}
{Erwin}, P., {Beckman}, J.~E., \& {Pohlen}, M. 2005, \apjl, 626, L81

\bibitem[{{Erwin} {et~al.}(2012){Erwin}, {Guti{\'e}rrez}, \&
  {Beckman}}]{Erwin2012}
{Erwin}, P., {Guti{\'e}rrez}, L., \& {Beckman}, J.~E. 2012, \apjl, 744, L11

\bibitem[{{Erwin} {et~al.}(2008){Erwin}, {Pohlen}, \& {Beckman}}]{Erwin:2008}
{Erwin}, P., {Pohlen}, M., \& {Beckman}, J.~E. 2008, \aj, 135, 20

\bibitem[{{Erwin} \& {Sparke}(2003)}]{Erwin2003}
{Erwin}, P. \& {Sparke}, L.~S. 2003, \apjs, 146, 299

\bibitem[{{Ferreira} {et~al.}(2022){Ferreira}, {Adams}, {Conselice},
  {Sazonova}, {Austin}, {Caruana}, {Ferrari}, {Verma}, {Trussler},
  {Broadhurst}, {Diego}, {Frye}, {Pascale}, {Wilkins}, {Windhorst}, \&
  {Zitrin}}]{Ferreira2022a}
{Ferreira}, L., {Adams}, N., {Conselice}, C.~J., {et~al.} 2022, \apjl, 938, L2

\bibitem[{{Ferreira} {et~al.}(2023){Ferreira}, {Conselice}, {Sazonova},
  {Ferrari}, {Caruana}, {Tohill}, {Lucatelli}, {Adams}, {Irodotou}, {Marshall},
  {Roper}, {Lovell}, {Verma}, {Austin}, {Trussler}, \&
  {Wilkins}}]{Ferreira2022b}
{Ferreira}, L., {Conselice}, C.~J., {Sazonova}, E., {et~al.} 2023, \apj, 955,
  94

\bibitem[{{Finkelstein} {et~al.}(2022){Finkelstein}, {Bagley}, {Haro},
  {Dickinson}, {Ferguson}, {Kartaltepe}, {Papovich}, {Burgarella}, {Kocevski},
  {Huertas-Company}, {Iyer}, {Koekemoer}, {Larson}, {P{\'e}rez-Gonz{\'a}lez},
  {Rose}, {Tacchella}, {Wilkins}, {Chworowsky}, {Medrano}, {Morales},
  {Somerville}, {Yung}, {Fontana}, {Giavalisco}, {Grazian}, {Grogin}, {Kewley},
  {Kirkpatrick}, {Kurczynski}, {Lotz}, {Pentericci}, {Pirzkal}, {Ravindranath},
  {Ryan}, {Trump}, {Yang}, {Almaini}, {Amor{\'\i}n}, {Annunziatella},
  {Backhaus}, {Barro}, {Behroozi}, {Bell}, {Bhatawdekar}, {Bisigello}, {Bromm},
  {Buat}, {Buitrago}, {Calabr{\`o}}, {Casey}, {Castellano}, {Ch{\'a}vez Ortiz},
  {Ciesla}, {Cleri}, {Cohen}, {Cole}, {Cooke}, {Cooper}, {Cooray}, {Costantin},
  {Cox}, {Croton}, {Daddi}, {Dav{\'e}}, {de La Vega}, {Dekel}, {Elbaz},
  {Estrada-Carpenter}, {Faber}, {Fern{\'a}ndez}, {Finkelstein}, {Freundlich},
  {Fujimoto}, {Garc{\'\i}a-Argum{\'a}nez}, {Gardner}, {Gawiser},
  {G{\'o}mez-Guijarro}, {Guo}, {Hamblin}, {Hamilton}, {Hathi}, {Holwerda},
  {Hirschmann}, {Hutchison}, {Jaskot}, {Jha}, {Jogee}, {Juneau}, {Jung},
  {Kassin}, {Bail}, {Leung}, {Lucas}, {Magnelli}, {Mantha}, {Matharu},
  {McGrath}, {McIntosh}, {Merlin}, {Mobasher}, {Newman}, {Nicholls}, {Pandya},
  {Rafelski}, {Ronayne}, {Santini}, {Seill{\'e}}, {Shah}, {Shen}, {Simons},
  {Snyder}, {Stanway}, {Straughn}, {Teplitz}, {Vanderhoof}, {Vega-Ferrero},
  {Wang}, {Weiner}, {Willmer}, {Wuyts}, {Zavala}, \& {Ceers
  Team}}]{Finkelstein:2022}
{Finkelstein}, S.~L., {Bagley}, M.~B., {Haro}, P.~A., {et~al.} 2022, \apjl,
  940, L55

\bibitem[{{Freeman}(1970)}]{freeman_1970}
{Freeman}, K.~C. 1970, \apj, 160, 811

\bibitem[{{Fudamoto} {et~al.}(2022){Fudamoto}, {Inoue}, \&
  {Sugahara}}]{Fudamoto2022}
{Fudamoto}, Y., {Inoue}, A.~K., \& {Sugahara}, Y. 2022, \apjl, 938, L24

\bibitem[{{Geach} {et~al.}(2011){Geach}, {Smail}, {Moran}, {MacArthur},
  {Lagos}, \& {Edge}}]{Geach2011}
{Geach}, J.~E., {Smail}, I., {Moran}, S.~M., {et~al.} 2011, \apjl, 730, L19

\bibitem[{{Gladders} \& {Yee}(2005)}]{Gladders2005}
{Gladders}, M.~D. \& {Yee}, H.~K.~C. 2005, \apjs, 157, 1

\bibitem[{{Guo} {et~al.}(2023){Guo}, {Jogee}, {Finkelstein}, {Chen}, {Wise},
  {Bagley}, {Barro}, {Wuyts}, {Kocevski}, {Kartaltepe}, {McGrath}, {Ferguson},
  {Mobasher}, {Giavalisco}, {Lucas}, {Zavala}, {Lotz}, {Grogin},
  {Huertas-Company}, {Vega-Ferrero}, {Hathi}, {Arrabal Haro}, {Dickinson},
  {Koekemoer}, {Papovich}, {Pirzkal}, {Yung}, {Backhaus}, {Bell},
  {Calabr{\`o}}, {Cleri}, {Coogan}, {Cooper}, {Costantin}, {Croton}, {Davis},
  {Dekel}, {Franco}, {Gardner}, {Holwerda}, {Hutchison}, {Pandya},
  {P{\'e}rez-Gonz{\'a}lez}, {Ravindranath}, {Rose}, {Trump}, {de la Vega}, \&
  {Wang}}]{Guo2023}
{Guo}, Y., {Jogee}, S., {Finkelstein}, S.~L., {et~al.} 2023, \apjl, 945, L10

\bibitem[{{Guti{\'e}rrez} {et~al.}(2011){Guti{\'e}rrez}, {Erwin}, {Aladro}, \&
  {Beckman}}]{Gutirrez_2011}
{Guti{\'e}rrez}, L., {Erwin}, P., {Aladro}, R., \& {Beckman}, J.~E. 2011, \aj,
  142, 145

\bibitem[{{Hasselfield} {et~al.}(2013){Hasselfield}, {Hilton}, {Marriage},
  {Addison}, {Barrientos}, {Battaglia}, {Battistelli}, {Bond}, {Crichton},
  {Das}, {Devlin}, {Dicker}, {Dunkley}, {D{\"u}nner}, {Fowler}, {Gralla},
  {Hajian}, {Halpern}, {Hincks}, {Hlozek}, {Hughes}, {Infante}, {Irwin},
  {Kosowsky}, {Marsden}, {Menanteau}, {Moodley}, {Niemack}, {Nolta}, {Page},
  {Partridge}, {Reese}, {Schmitt}, {Sehgal}, {Sherwin}, {Sievers}, {Sif{\'o}n},
  {Spergel}, {Staggs}, {Swetz}, {Switzer}, {Thornton}, {Trac}, \&
  {Wollack}}]{Hasselfield2013}
{Hasselfield}, M., {Hilton}, M., {Marriage}, T.~A., {et~al.} 2013, \jcap, 2013,
  008

\bibitem[{{Kartaltepe} {et~al.}(2023){Kartaltepe}, {Rose}, {Vanderhoof},
  {McGrath}, {Costantin}, {Cox}, {Yung}, {Kocevski}, {Wuyts}, {Ferguson},
  {Bagley}, {Finkelstein}, {Amor{\'\i}n}, {Andrews}, {Haro}, {Backhaus},
  {Behroozi}, {Bisigello}, {Calabr{\`o}}, {Casey}, {Coogan}, {Cooper},
  {Croton}, {de la Vega}, {Dickinson}, {Fontana}, {Franco}, {Grazian},
  {Grogin}, {Hathi}, {Holwerda}, {Huertas-Company}, {Iyer}, {Jogee}, {Jung},
  {Kewley}, {Kirkpatrick}, {Koekemoer}, {Liu}, {Lotz}, {Lucas}, {Newman},
  {Pacifici}, {Pandya}, {Papovich}, {Pentericci}, {P{\'e}rez-Gonz{\'a}lez},
  {Petersen}, {Pirzkal}, {Rafelski}, {Ravindranath}, {Simons}, {Snyder},
  {Somerville}, {Stanway}, {Straughn}, {Tacchella}, {Trump}, {Vega-Ferrero},
  {Wilkins}, {Yang}, \& {Zavala}}]{Kartaltepe2023}
{Kartaltepe}, J.~S., {Rose}, C., {Vanderhoof}, B.~N., {et~al.} 2023, \apjl,
  946, L15

\bibitem[{{Kennicutt}(1989)}]{Kennicutt1989}
{Kennicutt}, Robert~C., J. 1989, \apj, 344, 685

\bibitem[{{Laine} {et~al.}(2016){Laine}, {Laurikainen}, \& {Salo}}]{laine_2016}
{Laine}, J., {Laurikainen}, E., \& {Salo}, H. 2016, \aap, 596, A25

\bibitem[{{Laine} {et~al.}(2014){Laine}, {Laurikainen}, {Salo}, {Comer{\'o}n},
  {Buta}, {Zaritsky}, {Athanassoula}, {Bosma}, {Mu{\~n}oz-Mateos}, {Gadotti},
  {Hinz}, {Erroz-Ferrer}, {Gil de Paz}, {Kim}, {Men{\'e}ndez-Delmestre},
  {Mizusawa}, {Regan}, {Seibert}, \& {Sheth}}]{Laine_2014}
{Laine}, J., {Laurikainen}, E., {Salo}, H., {et~al.} 2014, \mnras, 441, 1992

\bibitem[{{Le Conte} {et~al.}(2023){Le Conte}, {Gadotti}, {Ferreira},
  {Conselice}, {de S{\'a}-Freitas}, {Kim}, {Neumann}, {Fragkoudi},
  {Athanassoula}, \& {Adams}}]{LeConte:2023}
{Le Conte}, Z.~A., {Gadotti}, D.~A., {Ferreira}, L., {et~al.} 2023, arXiv
  e-prints, arXiv:2309.10038

\bibitem[{{Liang} {et~al.}(2023){Liang}, {Yu}, {Fang}, \& {Ho}}]{Liang2023}
{Liang}, X., {Yu}, S.-Y., {Fang}, T., \& {Ho}, L.~C. 2023, arXiv e-prints,
  arXiv:2311.04019

\bibitem[{{Lin} \& {Pringle}(1987)}]{lin_prigles_1987}
{Lin}, D.~N.~C. \& {Pringle}, J.~E. 1987, \apjl, 320, L87

\bibitem[{{Lucy}(1974)}]{Lucy:1974}
{Lucy}, L.~B. 1974, \aj, 79, 745

\bibitem[{{Magdis} {et~al.}(2012){Magdis}, {Daddi}, {Sargent}, {Elbaz},
  {Gobat}, {Dannerbauer}, {Feruglio}, {Tan}, {Rigopoulou}, {Charmandaris},
  {Dickinson}, {Reddy}, \& {Aussel}}]{Magdis2012}
{Magdis}, G.~E., {Daddi}, E., {Sargent}, M., {et~al.} 2012, \apjl, 758, L9

\bibitem[{{Martin} \& {Kennicutt}(2001)}]{martin2001}
{Martin}, C.~L. \& {Kennicutt}, Robert~C., J. 2001, \apj, 555, 301

\bibitem[{{Mart{\'\i}n-Navarro} {et~al.}(2012){Mart{\'\i}n-Navarro}, {Bakos},
  {Trujillo}, {Knapen}, {Athanassoula}, {Bosma}, {Comer{\'o}n}, {Elmegreen},
  {Erroz-Ferrer}, {Gadotti}, {Gil de Paz}, {Hinz}, {Ho}, {Holwerda}, {Kim},
  {Laine}, {Laurikainen}, {Men{\'e}ndez-Delmestre}, {Mizusawa},
  {Mu{\~n}oz-Mateos}, {Regan}, {Salo}, {Seibert}, \& {Sheth}}]{Martin2012}
{Mart{\'\i}n-Navarro}, I., {Bakos}, J., {Trujillo}, I., {et~al.} 2012, \mnras,
  427, 1102

\bibitem[{{Mart{\'\i}n-Navarro} {et~al.}(2014){Mart{\'\i}n-Navarro},
  {Trujillo}, {Knapen}, {Bakos}, \& {Fliri}}]{Martin2014}
{Mart{\'\i}n-Navarro}, I., {Trujillo}, I., {Knapen}, J.~H., {Bakos}, J., \&
  {Fliri}, J. 2014, \mnras, 441, 2809

\bibitem[{{Mart{\'\i}nez-Bautista} {et~al.}(2021){Mart{\'\i}nez-Bautista},
  {Vel{\'a}zquez}, {P{\'e}rez-Villegas}, \& {Moreno}}]{Martinez-Bautista2021}
{Mart{\'\i}nez-Bautista}, G., {Vel{\'a}zquez}, H., {P{\'e}rez-Villegas}, A., \&
  {Moreno}, E. 2021, \mnras, 504, 5919

\bibitem[{{Miller} \& {van Dokkum}(2021)}]{Miller:2021}
{Miller}, T.~B. \& {van Dokkum}, P. 2021, \apj, 923, 124

\bibitem[{{Minchev} {et~al.}(2012){Minchev}, {Famaey}, {Quillen}, {Di Matteo},
  {Combes}, {Vlaji{\'c}}, {Erwin}, \& {Bland-Hawthorn}}]{Minchev2012}
{Minchev}, I., {Famaey}, B., {Quillen}, A.~C., {et~al.} 2012, \aap, 548, A126

\bibitem[{{Mortlock} {et~al.}(2013){Mortlock}, {Conselice}, {Hartley},
  {Ownsworth}, {Lani}, {Bluck}, {Almaini}, {Duncan}, {van der Wel},
  {Koekemoer}, {Dekel}, {Dav{\'e}}, {Ferguson}, {de Mello}, {Newman}, {Faber},
  {Grogin}, {Kocevski}, \& {Lai}}]{Mortlock2013}
{Mortlock}, A., {Conselice}, C.~J., {Hartley}, W.~G., {et~al.} 2013, \mnras,
  433, 1185

\bibitem[{{Mu{\~n}oz-Mateos} {et~al.}(2013){Mu{\~n}oz-Mateos}, {Sheth}, {Gil de
  Paz}, {Meidt}, {Athanassoula}, {Bosma}, {Comer{\'o}n}, {Elmegreen},
  {Elmegreen}, {Erroz-Ferrer}, {Gadotti}, {Hinz}, {Ho}, {Holwerda}, {Jarrett},
  {Kim}, {Knapen}, {Laine}, {Laurikainen}, {Madore}, {Menendez-Delmestre},
  {Mizusawa}, {Regan}, {Salo}, {Schinnerer}, {Seibert}, {Skibba}, \&
  {Zaritsky}}]{Munoz-Mateos2013}
{Mu{\~n}oz-Mateos}, J.~C., {Sheth}, K., {Gil de Paz}, A., {et~al.} 2013, \apj,
  771, 59

\bibitem[{{Pacaud} {et~al.}(2007){Pacaud}, {Pierre}, {Adami}, {Altieri},
  {Andreon}, {Chiappetti}, {Detal}, {Duc}, {Galaz}, {Gueguen}, {Le F{\`e}vre},
  {Hertling}, {Libbrecht}, {Melin}, {Ponman}, {Quintana}, {Refregier},
  {Sprimont}, {Surdej}, {Valtchanov}, {Willis}, {Alloin}, {Birkinshaw},
  {Bremer}, {Garcet}, {Jean}, {Jones}, {Le F{\`e}vre}, {Maccagni}, {Mazure},
  {Proust}, {R{\"o}ttgering}, \& {Trinchieri}}]{Pacaud2007}
{Pacaud}, F., {Pierre}, M., {Adami}, C., {et~al.} 2007, \mnras, 382, 1289

\bibitem[{{P{\'e}rez}(2004)}]{Perez2004}
{P{\'e}rez}, I. 2004, \aap, 427, L17

\bibitem[{{Perrin} {et~al.}(2014){Perrin}, {Sivaramakrishnan}, {Lajoie},
  {Elliott}, {Pueyo}, {Ravindranath}, \& {Albert}}]{Perrin:2014}
{Perrin}, M.~D., {Sivaramakrishnan}, A., {Lajoie}, C.-P., {et~al.} 2014, in
  Society of Photo-Optical Instrumentation Engineers (SPIE) Conference Series,
  Vol. 9143, Space Telescopes and Instrumentation 2014: Optical, Infrared, and
  Millimeter Wave, ed. J.~{Oschmann}, Jacobus~M., M.~{Clampin}, G.~G. {Fazio},
  \& H.~A. {MacEwen}, 91433X

\bibitem[{{Peters} {et~al.}(2017){Peters}, {van der Kruit}, {Knapen},
  {Trujillo}, {Fliri}, {Cisternas}, \& {Kelvin}}]{Peters2017}
{Peters}, S.~P.~C., {van der Kruit}, P.~C., {Knapen}, J.~H., {et~al.} 2017,
  \mnras, 470, 427

\bibitem[{{Pfeffer} {et~al.}(2022){Pfeffer}, {Bekki}, {Forbes}, {Couch}, \&
  {Koribalski}}]{Pfeffer2022}
{Pfeffer}, J.~L., {Bekki}, K., {Forbes}, D.~A., {Couch}, W.~J., \&
  {Koribalski}, B.~S. 2022, \mnras, 509, 261

\bibitem[{{Pohlen} \& {Trujillo}(2006)}]{pohlen2006}
{Pohlen}, M. \& {Trujillo}, I. 2006, \aap, 454, 759

\bibitem[{{Pranger} {et~al.}(2017){Pranger}, {Trujillo}, {Kelvin}, \&
  {Cebri{\'a}n}}]{Pranger2017}
{Pranger}, F., {Trujillo}, I., {Kelvin}, L.~S., \& {Cebri{\'a}n}, M. 2017,
  \mnras, 467, 2127

\bibitem[{{Raj} {et~al.}(2019){Raj}, {Iodice}, {Napolitano}, {Spavone}, {Su},
  {Peletier}, {Davis}, {Zabel}, {Hilker}, {Mieske}, {Falcon Barroso},
  {Cantiello}, {van de Ven}, {Watkins}, {Salo}, {Schipani}, {Capaccioli}, \&
  {Venhola}}]{Raj2019}
{Raj}, M.~A., {Iodice}, E., {Napolitano}, N.~R., {et~al.} 2019, \aap, 628, A4

\bibitem[{{Richardson}(1972)}]{Richardson:1972}
{Richardson}, W.~H. 1972, Journal of the Optical Society of America
  (1917-1983), 62, 55

\bibitem[{{Rodriguez-Gomez} {et~al.}(2015){Rodriguez-Gomez}, {Genel},
  {Vogelsberger}, {Sijacki}, {Pillepich}, {Sales}, {Torrey}, {Snyder},
  {Nelson}, {Springel}, {Ma}, \& {Hernquist}}]{Rodriguez-Gomez2015}
{Rodriguez-Gomez}, V., {Genel}, S., {Vogelsberger}, M., {et~al.} 2015, \mnras,
  449, 49

\bibitem[{{Roediger} {et~al.}(2012){Roediger}, {Courteau},
  {S{\'a}nchez-Bl{\'a}zquez}, \& {McDonald}}]{Roediger2012}
{Roediger}, J.~C., {Courteau}, S., {S{\'a}nchez-Bl{\'a}zquez}, P., \&
  {McDonald}, M. 2012, \apj, 758, 41

\bibitem[{{Ro{\v{s}}kar} {et~al.}(2008){Ro{\v{s}}kar}, {Debattista}, {Stinson},
  {Quinn}, {Kaufmann}, \& {Wadsley}}]{Roskar2008}
{Ro{\v{s}}kar}, R., {Debattista}, V.~P., {Stinson}, G.~S., {et~al.} 2008,
  \apjl, 675, L65

\bibitem[{{Schaye}(2004)}]{Schaye_2004}
{Schaye}, J. 2004, \apj, 609, 667

\bibitem[{{Sellwood} \& {Binney}(2002)}]{sellwood2002}
{Sellwood}, J.~A. \& {Binney}, J.~J. 2002, \mnras, 336, 785

\bibitem[{{Skelton} {et~al.}(2014){Skelton}, {Whitaker}, {Momcheva}, {Brammer},
  {van Dokkum}, {Labb{\'e}}, {Franx}, {van der Wel}, {Bezanson}, {Da Cunha},
  {Fumagalli}, {F{\"o}rster Schreiber}, {Kriek}, {Leja}, {Lundgren}, {Magee},
  {Marchesini}, {Maseda}, {Nelson}, {Oesch}, {Pacifici}, {Patel}, {Price},
  {Rix}, {Tal}, {Wake}, \& {Wuyts}}]{Skelton2014}
{Skelton}, R.~E., {Whitaker}, K.~E., {Momcheva}, I.~G., {et~al.} 2014, \apjs,
  214, 24

\bibitem[{{Tang} {et~al.}(2020){Tang}, {Chen}, {Zhang}, {Lin}, {Chen}, {Gao},
  {Liang}, {Liu}, \& {Kong}}]{tang2020}
{Tang}, Y., {Chen}, Q., {Zhang}, H.-X., {et~al.} 2020, \apj, 897, 79

\bibitem[{{Trujillo} \& {Fliri}(2016)}]{Trujillo:2016}
{Trujillo}, I. \& {Fliri}, J. 2016, \apj, 823, 123

\bibitem[{{Trujillo} \& {Pohlen}(2005)}]{Trujillo2005}
{Trujillo}, I. \& {Pohlen}, M. 2005, \apjl, 630, L17

\bibitem[{{Valentino} {et~al.}(2023){Valentino}, {Brammer}, {Gould}, {Kokorev},
  {Fujimoto}, {Jespersen}, {Vijayan}, {Weaver}, {Ito}, {Tanaka}, {Ilbert},
  {Magdis}, {Whitaker}, {Faisst}, {Gallazzi}, {Gillman}, {Gim{\'e}nez-Arteaga},
  {G{\'o}mez-Guijarro}, {Kubo}, {Heintz}, {Hirschmann}, {Oesch}, {Onodera},
  {Rizzo}, {Lee}, {Strait}, \& {Toft}}]{Valentino2023}
{Valentino}, F., {Brammer}, G., {Gould}, K. M.~L., {et~al.} 2023, \apj, 947, 20

\bibitem[{{van der Wel} {et~al.}(2014){van der Wel}, {Franx}, {van Dokkum},
  {Skelton}, {Momcheva}, {Whitaker}, {Brammer}, {Bell}, {Rix}, {Wuyts},
  {Ferguson}, {Holden}, {Barro}, {Koekemoer}, {Chang}, {McGrath},
  {H{\"a}ussler}, {Dekel}, {Behroozi}, {Fumagalli}, {Leja}, {Lundgren},
  {Maseda}, {Nelson}, {Wake}, {Patel}, {Labb{\'e}}, {Faber}, {Grogin}, \&
  {Kocevski}}]{vanderWel2014}
{van der Wel}, A., {Franx}, M., {van Dokkum}, P.~G., {et~al.} 2014, \apj, 788,
  28

\bibitem[{{Wang} {et~al.}(2018){Wang}, {Zheng}, {D'Souza}, {Mo}, {J{\'o}zsa},
  {Li}, {Kamphuis}, {Catinella}, {Shao}, {Lagos}, {Du}, \& {Pan}}]{wang:2018}
{Wang}, J., {Zheng}, Z., {D'Souza}, R., {et~al.} 2018, \mnras, 479, 4292

\bibitem[{{Watkins} {et~al.}(2019){Watkins}, {Laine}, {Comer{\'o}n}, {Janz}, \&
  {Salo}}]{watkins_2019}
{Watkins}, A.~E., {Laine}, J., {Comer{\'o}n}, S., {Janz}, J., \& {Salo}, H.
  2019, \aap, 625, A36

\bibitem[{{Williams} {et~al.}(2009){Williams}, {Quadri}, {Franx}, {van Dokkum},
  \& {Labb{\'e}}}]{Williams2009}
{Williams}, R.~J., {Quadri}, R.~F., {Franx}, M., {van Dokkum}, P., \&
  {Labb{\'e}}, I. 2009, \apj, 691, 1879

\bibitem[{{Yu} {et~al.}(2023){Yu}, {Cheng}, {Pan}, {Sun}, \& {Li}}]{Yu2023}
{Yu}, S.-Y., {Cheng}, C., {Pan}, Y., {Sun}, F., \& {Li}, Y.~A. 2023, \aap, 676,
  A74

\bibitem[{{Yu} {et~al.}(2022){Yu}, {Kalinova}, {Colombo}, {Bolatto}, {Wong},
  {Levy}, {Villanueva}, {S{\'a}nchez}, {Ho}, {Vogel}, {Teuben}, \&
  {Rubio}}]{Yu2022}
{Yu}, S.-Y., {Kalinova}, V., {Colombo}, D., {et~al.} 2022, \aap, 666, A175

\end{thebibliography}

\end{document}